\documentclass{article}

\RequirePackage{amsthm,amsmath,amsfonts,amssymb, bm}
\RequirePackage[authoryear]{natbib}
\usepackage{setspace}
\usepackage{imsart}
\usepackage{hyperref}
\usepackage{dsfont}
\RequirePackage[colorlinks,citecolor=blue,urlcolor=blue]{hyperref}
\RequirePackage{graphicx}
\usepackage{fullpage}

\begin{document}

\title{Fusing Climate Data Products using a Spatially Varying Autoencoder}
\author{Jacob~A.~Johnson, Matthew~J.~Heaton, William~F.~Christensen, \\ Lynsie~R.~Warr, and Summer~B.~Rupper}

\begin{abstract}
Autoencoders are powerful machine learning models used to compress information from multiple data sources.  However, autoencoders, like all artificial neural networks, are often unidentifiable and uninterpretable.  This research focuses on creating an identifiable and interpretable autoencoder that can be used to meld and combine climate data products.  The proposed autoencoder utilizes a Bayesian statistical framework, allowing for probabilistic interpretations while also varying spatially to capture useful spatial patterns across the various data products.  Constraints are placed on the autoencoder as it learns patterns in the data, creating an interpretable consensus that includes the important features from each input.  We demonstrate the utility of the autoencoder by combining information from multiple precipitation products in High Mountain Asia.
\end{abstract}
\maketitle


\section{Introduction}\label{intro}

\subsection{Problem Background}
\begin{figure}
\begin{center}
\includegraphics[scale=0.45]{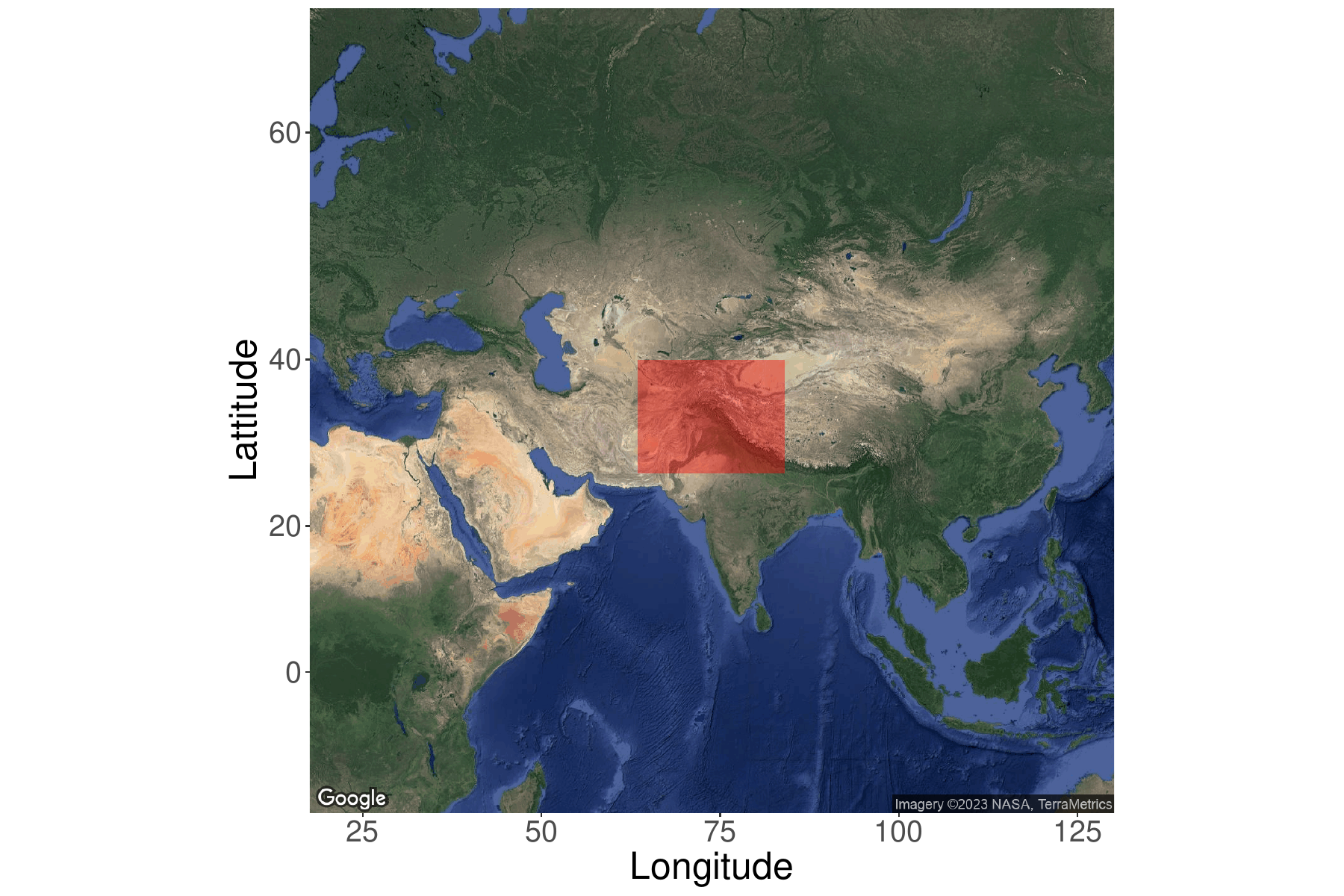}
\end{center}
\caption{World map showing the HMA region.}
\label{hma_map}
\end{figure}

Nearly 650 glaciers located throughout the high mountain Asia (HMA) region shown in Figure \ref{hma_map} \citep{ggmap} feed water into the Indus, Ganges, and Brahmaputra rivers.  These rivers, in turn, provide water resources for hundreds of millions of people \citep{zhang2019water}.  Given the high population in this region, understanding changes to precipitation and water availability in HMA is important for sustaining life.  Given the importance of the region, the National Aeronautics and Space Administration (NASA) coordinates the efforts of a group of research teams to understand climatological changes in the area (see \url{https://www.himat.org/}).

Given the highly mountainous terrain of this region, \textit{in situ}  measurements of precipitation via rain gauges are sparse at best \citep{maussion2014precipitation} and can also be unreliable \citep{palazzi2013precipitation}.  Hence, scientific understanding of precipitation in the region relies primarily on digital data products.  ``Digital data'' for HMA includes, but is not limited to, multi-scale computer simulations such as global and regional climate models \citep{grogan2022water}, remotely sensed data or even observation-based gridded data products such as reanalysis data \citep{stillinger2023landsat, warr2023distributional}.  Many of the digital data products for HMA are publicly available at \url{https://www.himat.org/gmelt/data/}.  

\begin{figure}
\begin{center}
\includegraphics[scale=0.45]{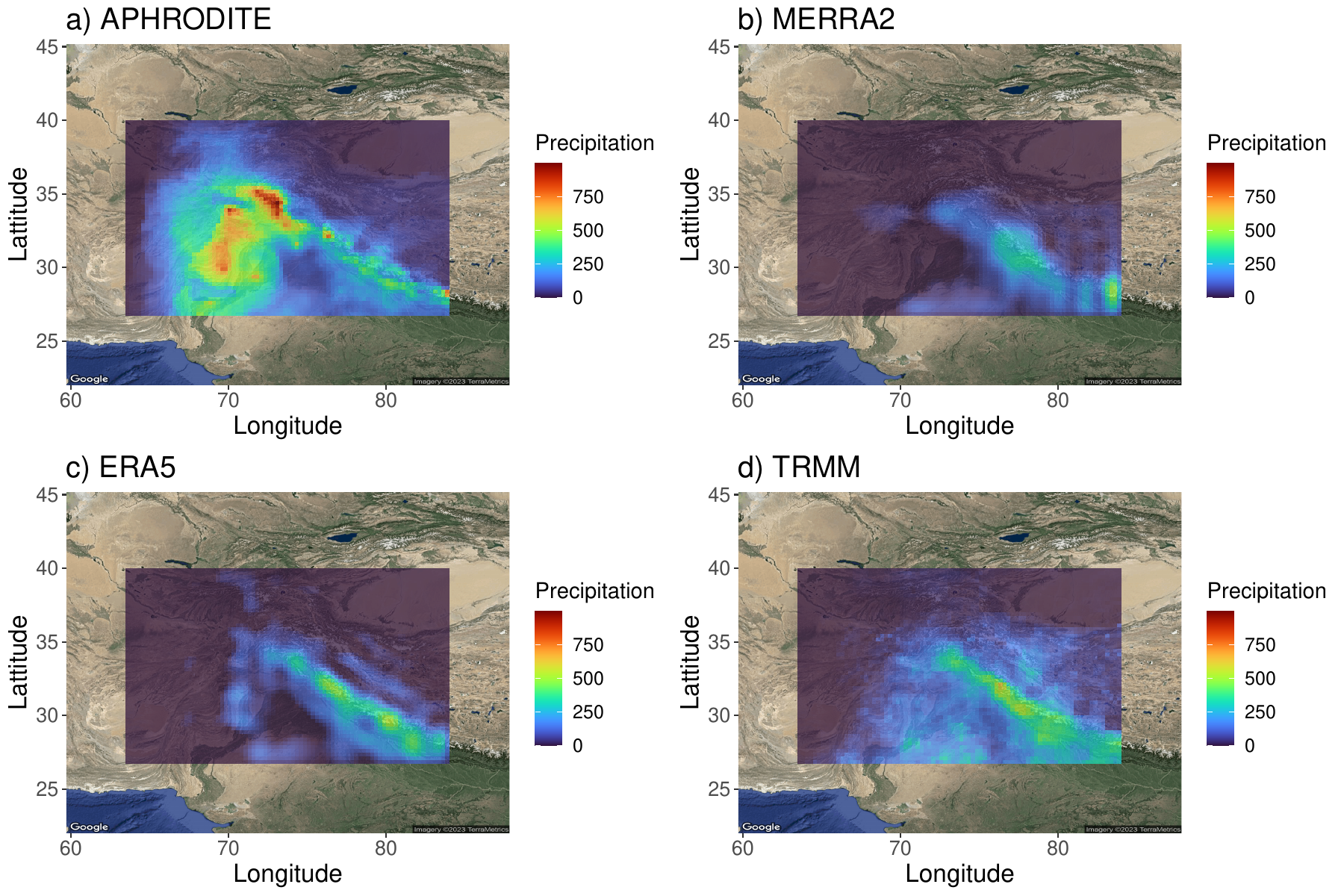}
\end{center}
\caption{Four different digital precipitation data products for July 2015.}
\label{map_data}
\end{figure}

One challenge of dealing with digital data products is that the various products commonly disagree in their measurements of climatological variables.  As an example, consider Figure \ref{map_data} which displays four different digital data products of precipitation in HMA.  Geographically, it is important to note that along the southern edge of the Tibetan plateau lies a broad escarpment where elevation changes dramatically over a relatively short distance.  This escarpment runs essentially diagonally in the southeast direction from $70^\circ$ longitude to the bottom right corner of the region.  This geological feature is important because precipitation typically gets halted by the mountains in this area.  Thus, the area to the north and east of this escarpment is a high elevation plateau with low precipitation compared to the relatively low elevation and high precipitation farmland to the south and west of the plateau.

Discrepancies between these digital data products often arise because different data products are built in unique ways for their specific scientific purposes.  For example, the Asian Precipitation --- Highly Resolved Observational Data Integration Towards Evaluation (APHRODITE) data product relies on statistical interpolation of rain gauge data and is generally used for detailed precipitation climatology as well as extreme events.  In contrast, the Modern-Era Retrospective analysis for Research and Applications, Version 2 (MERRA-2) utilizes data assimilation techniques \citep{carrassi2018data} and its ultimate goal is to provide an entire earth-system reanalysis.  The European Centre for Medium-Range Weather Forecasts Reanalysis v5 (ERA5) data is also global in nature (hence the lower resolution over HMA) and relies on weather forecasting models.  Finally, Tropical Rainfall Measuring Mission (TRMM) data is entirely remote sensing data and focuses on tropical area precipitation and, thus, may be less reliable over HMA than the tropical regions for which it was built.

Due to the differences in digital data products, a common statistical goal is to merge various data products into a single consensus data product through a process called data fusion (also known as data melding).  While articles and books such as \citet{schmitt2016data} and \citet{cocchi2019data} give a nice overview of data fusion techinques commonly used in atmospheric sciences, for purposes of highlighting the contribution of the methods proposed here, we highlight just a few examples from the statistical literature.  \citet{nguyen2012spatial} and \citet{christensen2019bayesian} represent each individual data product as being noisy representations of a spatially smooth fused data product (hereafter referred to as the ``consensus'').  While this is an intuitive setup, it has a few drawbacks.  First, the estimated consensus data product can be shown to be a linear combination of the individual data products which may not be realistic given the complexity of the relationship between the different data products.  And, second, due to the large size of the data products, spatial smoothing of the consensus product is done through basis function representations of a Gaussian process which have been shown to oversmooth the resulting product \citep{stein2014limitations}.

Both \citet{berrocal2010spatio} and \citet{neeley2014bayesian} posit a more complex relationship between the data products than those proposed above.  \citet{berrocal2010spatio} use spatially varying linear regression models while \citet{neeley2014bayesian} use spatially varying factor models.  The spatially-varying nature of their models allows for a more rich structure in the consensus but, ultimately, the consensus is still a linear combination of the data products.  Further, the spatially-varying nature of the model vastly complicates the computation needed for fitting.

In more recent years, machine and deep learning have become commonplace for data fusion.  \citet{meng2020survey} and \citet{li2022deep} provide nice reviews of approaches utilizing machine or deep learning but approaches span the breadth of tools including random forests \citep{zhang2020fusion}, support vector machines \citep{challa2002distributed} and neural networks \citep{starzacher2009embedded,yin2020spatiotemporal,jia2022multitask}.  The biggest issue with deep or machine learning approaches is the lack of uncertainty quantification in the consensus data product \citep{DUNSON20184}. To this end, many have begun using deep learning embedded within statistical modeling so as to have some measure of uncertainty \citep{wikle2023statistical} but little in this area has been done in terms of data fusion.  Even with the hybrid approach however, deep or machine learning models still lack in terms of interpretability of model parameters.

The primary contribution of this research is to propose an autoencoder to fuse the precipitation data products shown above.  An autoencoder \citep[see][]{zhai2018autoencoder} is a neural network architecture commonly used for dimension reduction that can be viewed as a non-linear extension of principal components \citep{ladjal2019pca}.  Specifically, we utilize the autoencoder architecture to achieve non-linear data fusion.  Further, we impose identifiability constraints on the autoencoder which allows the resulting consensus product to be interpretable.  To capture a rich spatial structure in the data fusion, we allow the weights and biases of the auto encoder to vary over space.  Finally, to capture uncertainty in the resulting consensus data product, we estimate the autoencoder parameters using the Bayesian paradigm.

The remainder of this paper is outlined as follows.  Section \ref{mod} provides the details of the spatially-varying autoencoder along with a discussion of the constraints needed to ensure the consensus is interpretable as a data product.  Section \ref{app} then fits the autoencoder to the precipitation data shown in Figure \ref{map_data}.  Finally, Section \ref{conc} draws conclusions and points out areas of future research.

\section{Statistical Model}\label{mod}

\subsection{Model Structure}
Let $Y_d(\bm{s})$ represent the response from data product $d = 1,\dots,D$ at location $\bm{s} \in \Omega$ where $\Omega$ is the spatial domain (in our HMA application $Y_d(\bm{s})$ is the precipitation data displayed in Figure \ref{map_data}).  Generally, we let
\begin{align}
    Y_d(\bm{s}) = g(Z_d(\bm{s}))
\end{align}
where $g(\cdot)$ is some transformation function that preserves the support of the data and $Z_d(\bm{s})$ is a latent variable.  For example, in our HMA precipitation application we use the rectified linear unit (RELU) transformation $g(z) = \max(z, 0)$ because $Y_d(\bm{s}) \geq 0$ such that $Z_d(\bm{s}) = Y_d(\bm{s})$ if $Z_d(\bm{s})>0$ but the choice of $g(\cdot)$ will be application specific.  Other possible choices for $g(\cdot)$ include the identity transformation (for quantitative data) or, for binary data, a cutoff transformation such as $g(z) = \mathds{1}(z>c)$ where $\mathds{1}(\cdot)$ is an indicator function.  

\begin{figure}[tb]
\includegraphics[scale=0.5]{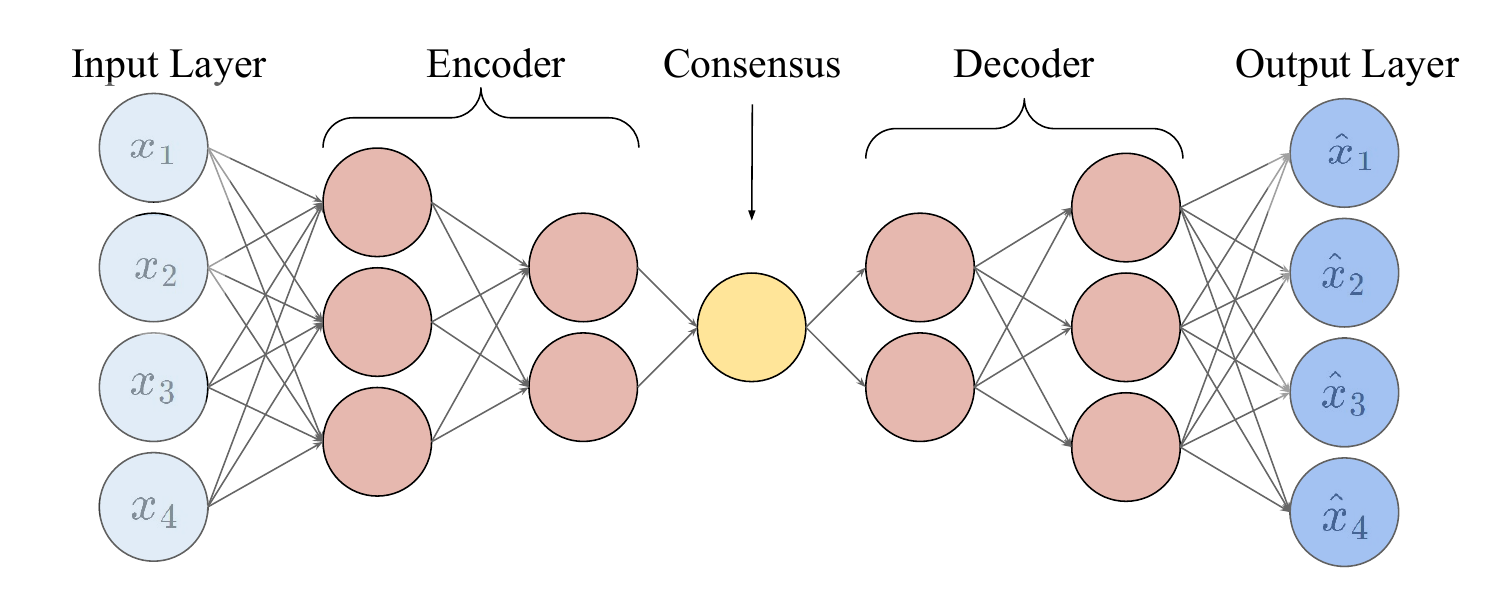}
\caption{Autoencoder structure with a single neuron consensus.}
\label{autoencoder}
\end{figure}

Because uncertainty quantification for our fused data product is of high importance, we specify a model for $Z_d(\bm{s})$.  Specifically, we assume,
\begin{align}
    Z_{d}(\bm{s}) &\overset{iid}{\sim} \mathcal{N}(N_{d}^{\text{out}}(\bm{s}),\sigma^2)
    \label{data_layer}
\end{align}
where $N_{d}^{\text{out}}(\bm{s})$ is the output layer of an autoencoder at location $\bm{s}$.  An autoencoder \citep{zhai2018autoencoder} is a unique neural network architecture specifically suited for dimension reduction and is illustrated by Figure \ref{autoencoder}.  Under an autoencoder, there are an equal number of inputs and outputs, which each output being a modified version of its corresponding input.  This modification occurs as the autoencoder non-linearly compresses the inputs into a lower dimensional space (a process referred to as ``encoding'') then subsequently decompresses this into the original input (a process referred to as ``decoding'').  Due to the dimension reduction, the output will not perfectly reproduce the input but the consensus neurons should be an optimal non-linear compression of the inputs that can best reproduce the inputs.  We note that while Figure \ref{autoencoder} displays a single neuron in the center, autoencoders generally can have more than one center neuron but we focus on the single neuron case here because we wish to compress the data into a single consensus product.

We begin to formally define our autoencoder, which has $P_\ell$ inputs at layer $\ell$, where $\ell=1,\dots,L$ is the index of the $L$ hidden layers of the autoencoder, by letting $\bm{N}^\ell(\bm{s})$ be the $P_\ell$ vector of neurons at layer $\ell$ defined by
\begin{align}
    \bm{N}^\ell(\bm{s}) &= a_\ell\left(\bm{w}_0^\ell(\bm{s}) + \bm{W}^\ell(\bm{s})\bm{N}^{\ell-1}(\bm{s})\right)
    \label{auto}
\end{align}
where $a_\ell(\cdot)$ is an activation function for layer $\ell$, $\bm{w}_0^\ell$ is a $P_\ell$ vector of biases, and $\bm{W}^\ell(\bm{s})$ is a $P_\ell \times P_{\ell-1}$ matrix of weights for transitioning from layer $\ell-1$ to layer $\ell$.  For completeness, the output layer is given by
\begin{align}
    \bm{N}^{\text{out}}(\bm{s}) &= a\left(\bm{w}_0^{\text{out}}(\bm{s}) + \bm{W}^{\text{out}}(\bm{s})\bm{N}^{L}(\bm{s})\right)\label{output_layer}
\end{align}
and we assume that $\bm{N}^{0}(\bm{s}) = (Y_1(\bm{s}),\dots,Y_d(\bm{s}))'$ are the $P_0$ input variables.  This model structure has several features that warrant further discussion, which we consider below.  

From \eqref{data_layer}, we are assuming spatial independence.  Given the motivating application, this may seem like a strong assumption.  However, recall that the autoencoder should be reproducing the inputs at the output layer.  Because the inputs have spatial structure, by construction, so should the output layer.  Intuitively, this is assuming that the spatial structure observed in each data product is captured by the autoencoder output.  This assumption is commonly made for data fusion problems as a way of enforcing smoothness on the consensus data product \citep[see][]{berrocal2010spatio}.

For autoencoders, there are generally no restrictions on the weights and biases in \eqref{auto} and \eqref{output_layer}.  Because of this, the consensus layer (neuron) is not generally interpretable.  However, in this research, interpretability of the consensus product is paramount.  To this end, we constrain the autoencoder parameters in two ways as follows.  First, we set $\bm{w}_0^\ell \equiv \bm{0}$ for all $l$ (including the output).  In this way, the autoencoder resembles the same structure as a principal components dimension reduction, except that the autoencoder allows for non-linear transformations through the activation functions.  

Second, note that the decoder portion of the autoencoder is akin to a latent factor model.  Leaning on the lower triangular constraints of confirmatory factor analysis \citep[see][]{renchermethods, neeley2014bayesian} to ensure identifiability of latent factors, we assume a lower triangular structure for each weight matrix in the decoder portion of the autoencoder.  For example, under the architecture in Figure \ref{autoencoder}, we would constrain
\begin{align}
    \bm{W}^{\text{out}}(s) &=  
    \begin{pmatrix}
        1 & 0 & 0 \\
        W_{2,1}^{\text{out}}(\bm{s}) & 1 & 0 \\
        W_{3,1}^{\text{out}}(\bm{s}) & W_{3,2}^{\text{out}}(\bm{s}) & 1 \\
        W_{4,1}^{\text{out}}(\bm{s}) & W_{4,2}^{\text{out}}(\bm{s}) & W_{4,3}^{\text{out}}(\bm{s})
    \end{pmatrix}
    \label{lower}
\end{align}
where $W_{p,j}^{\text{out}}(\bm{s})$ is the weight of the $j^{th}$ neuron at layer $L$ in creating the $p^{th}$ neuron at the output layer.  As discussed by \citet{neeley2014bayesian}, this constraint serves to tie down the scale of the consensus to match one of the data products - in this case the first product - without constraining the autoencoder to reproduce the output directly.

Finally, we note that each of the autoencoder parameters is location-specific (varies by location).  Even with the zero biases and lower-triangular constraints imposed on the decoder (see \eqref{lower} as an example), this results in a massively overparameterized model.  Under the autoencoder structure in Figure \ref{autoencoder}, there are 30 parameters (accounting for the constraints above) per location.  Given the size of the domain in Figure \ref{map_data}, this would equate to $30\times 4346 = 130,380$ parameters for only $4 \times 4346  = 17,384$ data points.  Hence, we enforce spatial structure in the weights to reduce the dimensionality as follows.

We model the weights (including the output layer weights) in \eqref{auto} as,
\begin{align}
    W_{p,j}^{l}(s) &= \bm{b}'(\bm{s})\bm{\theta}^{l}_{p,j}
    \label{spatbasis}
\end{align}
where $\bm{\theta}_{p,j}^l$ is a $K+1$ vector of coefficients and $\bm{b}(\bm{s}) = (1,b_1(\bm{s}),\dots,b_K(\bm{s}))'$ denote a set of $K$ spatial basis functions such as kernels, predictive processes or bisquare functions \citep[see][]{cressie2022basis}.  This basis function approach to spatial smoothing has a long successful history of modeling complex spatial functions \citep[see][]{cressie2022basis}.  Beyond capturing spatial heterogeneity in the autoencoder, this basis function approach reduces the parameter space of the weights to $30\times K$ parameters.  

While there are many choices of basis functions, we opt to use a bivariate Gaussian kernel with $K$ centering locations spread out as a grid over the domain in Figure \ref{map_data} (see \citealt{higdon1998process}).  As will be detailed in Section \ref{app} below, we fix the parameters of the Gaussian kernel but we consider the number of basis functions ($K$) as an additional tuning parameter in the model structure.

As a final note, we point out that this autoencoder is \textit{not} a generative model because the inputs to the autoencoder are also the outputs.  Hence, in a deviation from traditional statistical models, we use this autoencoder to do data fusion but it cannot be interpreted as a model for some larger population.  This makes our approach more akin to, say, principal components dimension reduction than the approaches of \citet{nguyen2012spatial} or \citet{christensen2019bayesian} which view data products as randomly arising from some population of data products.

\subsection{Model Fitting and Tuning}
\label{modfit}
Autoencoders (and neural networks more generally) are usually fit via loss minimization with backpropagation used as an efficient way to calculate the gradient of the loss function.  However, these fitting algorithms generally don't lend themselves to uncertainty quantification in the model fit.  While there are ad-hoc methods for quantifying uncertainty in neural networks \citep[see][]{kabir2018neural}, there is an ever-growing interest on the use of Bayesian methods for fitting neural networks because of their natural ability to quantify uncertainty \citep{polson2017deep, wikle2023statistical}.  Because uncertainty in the consensus product is important for our motivating application, we also adopt the Bayesian paradigm for estimating model parameters.

The unknown parameters of the above autoencoder are the weight coefficients $\{\bm{\theta}_{p,j}^l\}_{l,j,p}$ and, for our precipitation application, the variance $\sigma^2$ in \eqref{data_layer}.  We use fairly non-informative, independent $\mathcal{N}(0, 5)$ priors for the coefficients and an inverse-gamma prior with shape 2.1 and rate 1.1 for $\sigma^2$.  The Gaussian prior for the coefficients acts similar to an $L^2$ penalty which is commonly used as a tuning parameter in neural network models.  As we show below, we found that the resulting model fit was far more sensitive to the autoencoder architecture rather than the choice of priors.

Given the above priors, inference is carried out via Markov chain Monte Carlo (MCMC) sampling from the posterior distribution.  We use adaptive Metropolis steps \citep{haario2001adaptive} to update the $\bm{\theta}$ coefficients.  We found that updating all the coefficients associated with a given layer improved mixing of the chain and decreased computation time because more parameters are being updated simultaneously.  Specific to the HMA application, due to the truncation at 0 for observed precipitation, any $Z_d(\bm{s})$ for which $Y_d(\bm{s})=0$ is sampled directly from its negatively-truncated Gaussian complete conditional distribution.  Finally, the complete conditional for $\sigma^2$ is conjugate and updated directly.  Code for fitting our autoencoder is available at \url{https://github.com/jajcelloplayer/Weather-Autoencoder}.

Beyond model fitting, there are several tuning parameters associated with the autoencoder.  Namely, tuning parameters include the depth (number of layers), width of each layer and the number of basis functions used for spatial smoothing of the coefficients.  As a general strategy for tuning, we fit the autoencoder at various values of the tuning parameters and evaluate the performance using total root mean square error across the data products.  
Because we wish to achieve dimension reduction, we restrict our search of the number of hidden layers ($L$) to at most $2D-3$ where $D$ is the number of data products we wish to fuse and a maximum of $D$ neurons per layer with a single neuron at the center.  This ensures that the autoencoder structure resembles that shown in Figure \ref{autoencoder} but may have more neurons in the off-center layers.

\section{Application to HMA Data}\label{app}

\subsection{Setup and Model Fit}
Using root mean square error (RMSE) as the tuning criterion, the optimal autoencoder architecture had $L=5$ hidden layers with the number of neurons shown in Figure \ref{autoencoder}.  Further, we used $K=36$ bivariate normal kernels as our spatial basis functions in \eqref{spatbasis} with equally spaced centers on a $6\times6$ grid over the spatial domain.  As noted above, we opted to use a bivariate Gaussian kernel with covariance matrix $\text{diag}(1.9291^2, 1.255^2)$.  These values were chosen to evenly cover the spatial domain and were calculated by dividing the range of the data in each direction by the number of support points in that direction plus 1.  

To fit the autoencoder to the HMA data in Figure \ref{map_data}, we used a burn in of 100000 with an additional 50000 MCMC draws used for posterior inference.  We found slow mixing in the resulting Markov chain but it eventually converged after the 100000 burn-in iterations as evidenced by traceplots of the $-2\times$log-likelihood.  This is consistent with other work on Bayesian fitting of neural networks by \citet{polson2017deep} or \citet{d2022underspecification} in which neural network training parameters are only weakly identifiable.  While we hypothesize that the constraints we use in the autoencoder help with the identifiability of model parameters, we note that continued work may be needed to improve Bayesian fitting of neural networks.

In terms of model fit, we calculated a pseudo-$R^2$ of the output of the autocoder with each data product as $\text{Corr}(Y_d(\bm{s}), \hat{Y}_d(\bm{s}))^2$ where $\hat{Y}_d(\bm{s}) = g(\hat{Z}_d(\bm{s}))$ and $\hat{Z}_d(\bm{s})$ is the posterior mean of the $d^{th}$ output from the autoencoder.  These pseudo-$R^2$ values were $0.986, 0.921, 0.950$ and $0.947$ for APRHODITE, MERRA2, ERA5 and TRMM, respectively.  Given these high values of $R^2$, the autoencoder sufficiently fits the HMA data.  Further, the posterior mean of the autoencoder gave RMSEs of  29.28, 27.60, 27.84 and 34.99 for APRHODITE, MERRA2, ERA5 and TRMM respectively.

\subsection{Fused Data Product}

\begin{figure}
\begin{center}
\includegraphics[scale=0.4]{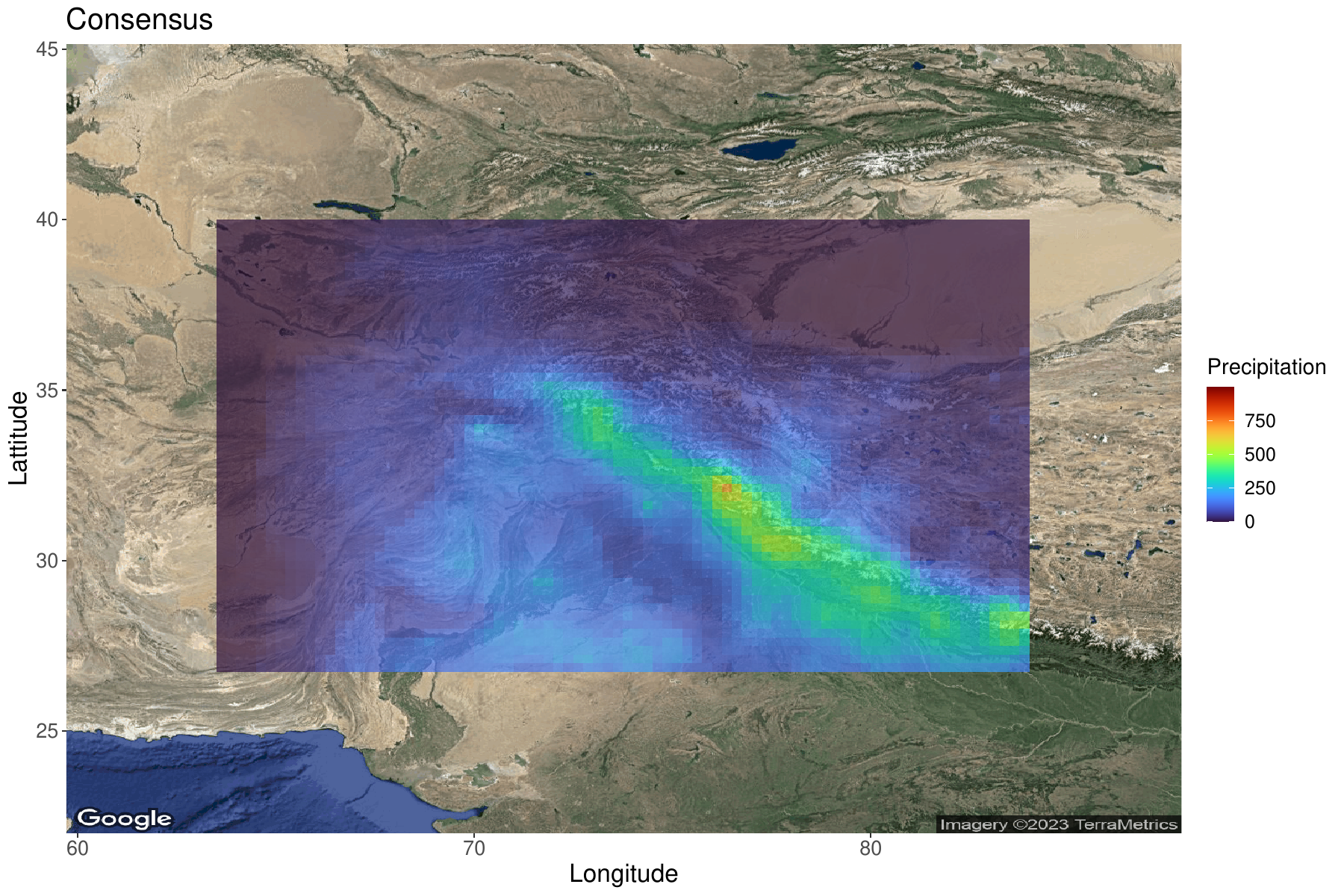}
\caption{Consensus Data Product for July 2015.}
\label{consensus_map}
\end{center}
\end{figure}

Figure \ref{consensus_map} shows the posterior mean of the autoencoded, fused data product from the July 2015 HMA data shown in Figure \ref{map_data}.  For this paper, we focus our discussion on the July 2015 HMA data but note that we successfully fit the autoencoder to over 20 years of monthly precipitation data from the same region from January 1998 to August 2018.  Images and a video for the fused data product across these 20 years of data is available at \url{https://github.com/jajcelloplayer/Weather-Autoencoder}.

As intended, the consensus product captures unique features of each data product.  Specifically, comparing Figure \ref{map_data} with Figure \ref{consensus_map}, the consensus has larger precipitation values in the west from APHRODITE but these are tempered because these high precipitation values are not observed in the other data products.  Further, the consensus captures the higher precipitation values from ERA5 and MERRA2 along the southern edge of the Tibetan plateau, while also closely following the TRMM data to the north-east of the plateau.

\begin{figure}
\begin{center}
\includegraphics[scale=0.45]{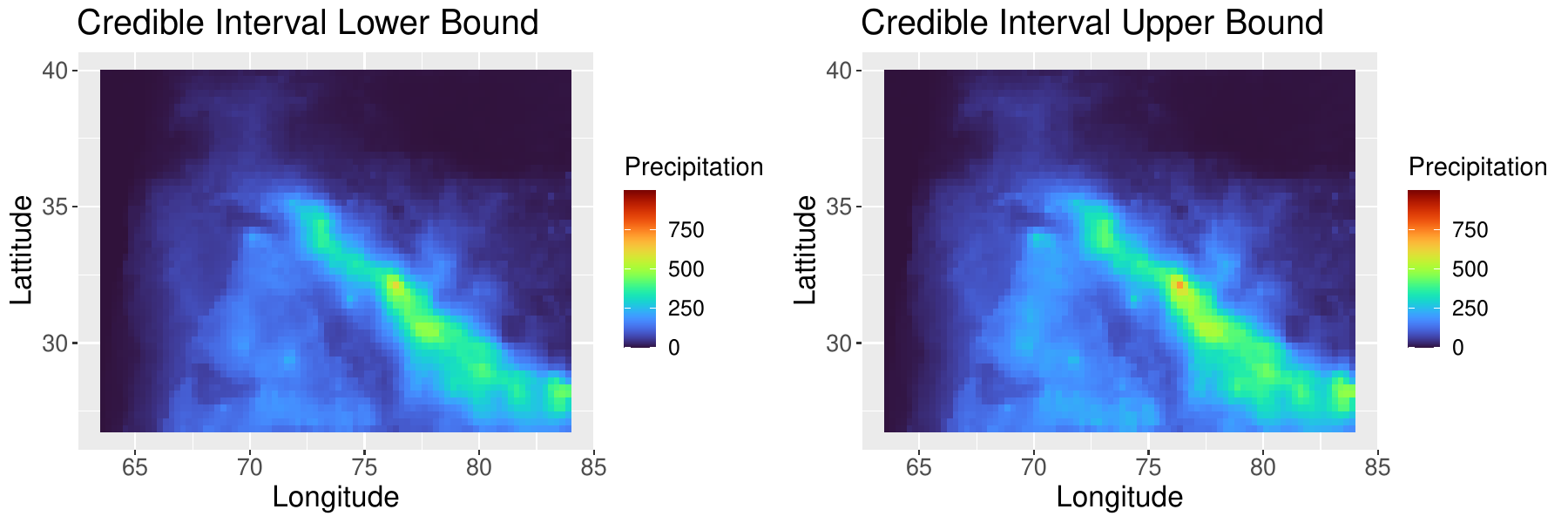}
\end{center}
\caption{95\% Credible Interval on Autoencoder Consensus.}
\label{uncertainty}
\end{figure}

As mentioned in \ref{modfit}, a major advantage to using Bayesian methods is the inclusion of uncertainty quantification about model parameters and predictions.  Using the saved MCMC draws, we can easily construct credible intervals and calculate probabilities of interest on the merged consensus.  The upper and lower bounds of a centered 95\% credible interval on the consensus are shown in \ref{uncertainty}.  Importantly, both the upper and lower bounds of this interval show the same spatial patterns as that seen in Figure \ref{consensus_map}.  

\begin{figure}
\begin{center}
\includegraphics[scale=0.45]{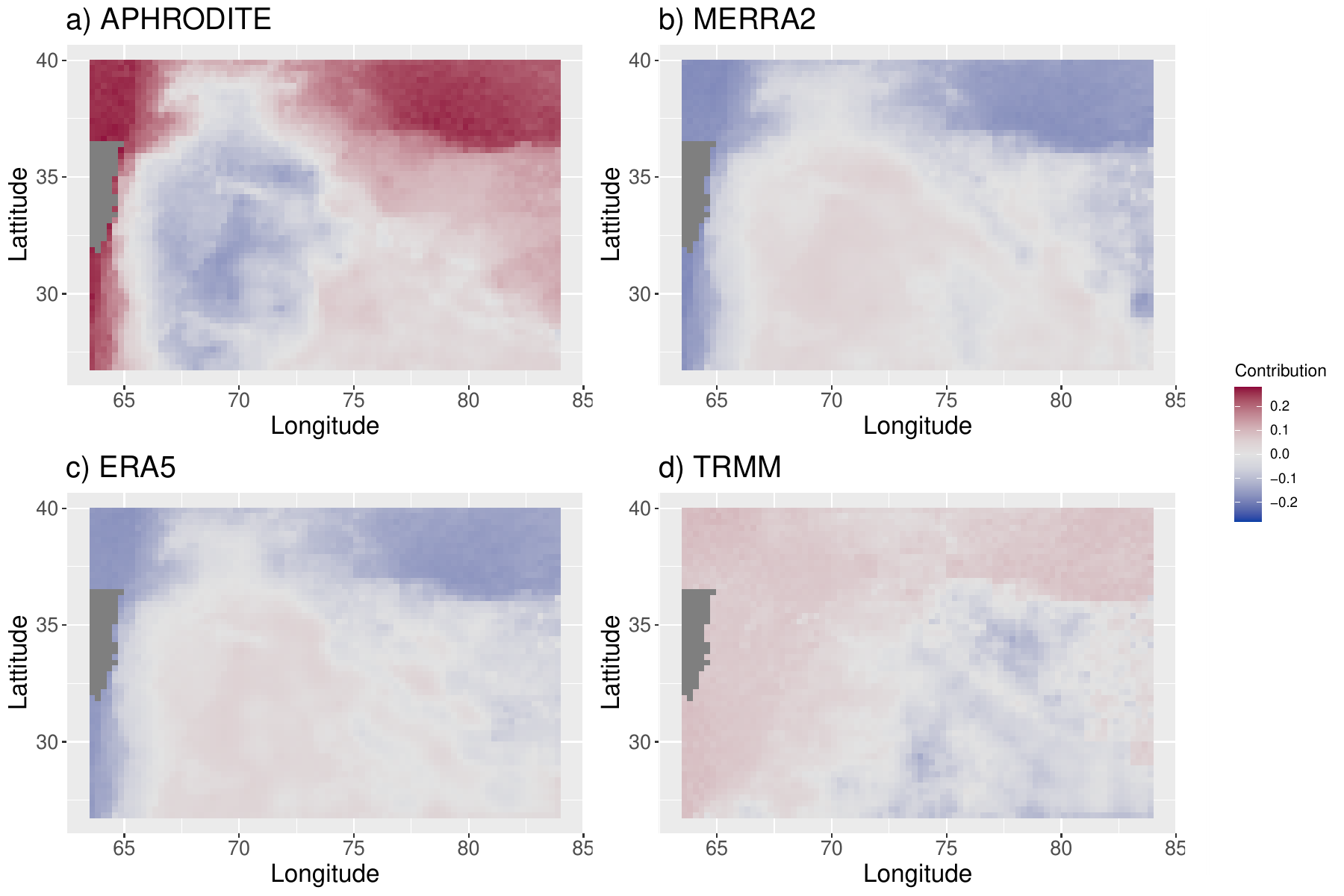}
\end{center}
\caption{Spatial Contribution of each Data Product.}
\label{contribution}
\end{figure}

To determine how each data product is contributing to the consensus in Figure \ref{consensus_map}, we plot the contribution of each data product in Figure \ref{contribution} (note that the contribution of each data product varies over space because the autoencoder weights vary over space).  This contribution metric was created using permutation importance \citep{molnar2022}.  Keeping all else the same, the precipitation values were shuffled for a given data product and the effect that this shuffling had on the consensus was measured.  The ``contribution'' or importance of each data product was normalized and centered in Figure \ref{contribution} so that a value of 0 represents a one-fourth contribution (equivalent to an equal weighting of all 4 data products in the area).  Further, positive (negative) values indicate that the data product contributed more (less) to the consensus than equal weighting.

Interestingly, all four data products contribute equally in the regions where they have similar values, while data products contribute less where they contain precipitation values that are distinct from the other data products.  In this way, the autoencoder penalizes areas where inputs are different from others.  For example, APRHODITE is downweighted in the western portion of the domain while TRMM is downweighted in the southeastern portion.  MERRA2 and ERA5 contribute similarly acoross sapce, and are most important in areas where APHRODITE and TRMM were downweighted for being different from all the others.  

Notably, Figure \ref{contribution} show that each data product has areas of above and below average contribution.  This spatial information would be lost by only looking at a single summary of the importance or contribution of each product.  The summary importance of each data product is 0.25, -0.17, -0.16, and 0.08 for APHRODITE, MERRA2, ERA5, and TRMM respectively.  From this summary, one may conclude that APHRODITE contributes more than average, yet there is a large region in which APHRODITE actually contributes considerably less than average, which is shown in blue in Figure \ref{contribution}a.  Thus, the autoencoder proposed here adds unique spatial information about how the consensus is created at each spatial location.

\subsection{Sensitivity and Comparison with Alternative Methods}
To investigate the sensitivity of the consensus data product to the lower triangular constraints, we fit the autoencoder to all possible orderings of the data products so that the constraints shown in Equation \eqref{lower} tie down the consensus to a different data product under these different orderings.  Regardless of the ordering of the data products, the spatial structure of the consensus was effectively the same as that shown in Figure \ref{consensus_map}.  However, the scale of the consensus varied to match the first data product in the ordering.  This is to be expected, however, because the weights for creating the first data product from the consensus in the decoder are all constrained to be 1.  Thus, is terms of climate science, practioners need to consider which product ultimately defines the scale of the fused data product.  For purposes of our application, we chose the TRMM data product to be the ``first'' data product because it was based entirely on remote sensing data but, certainly, another data product could be used.  

\begin{figure}
\begin{center}
\includegraphics[scale=0.45]{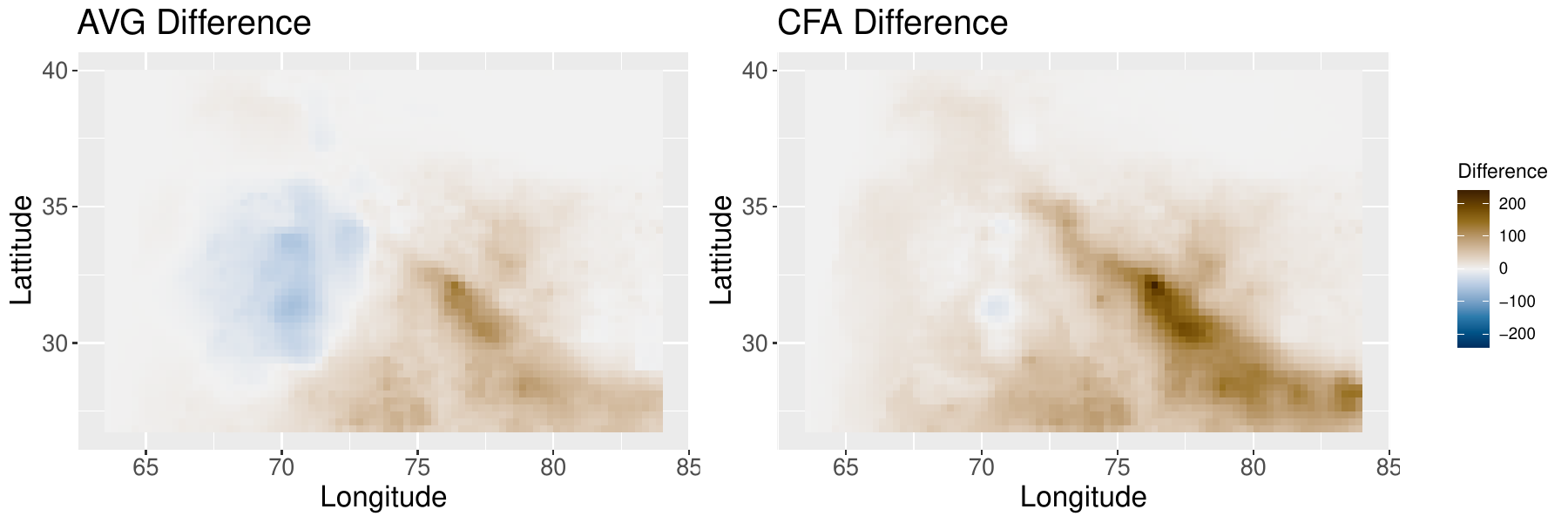}
\end{center}
\caption{Differences in Fused Consensus Data Products}
\label{differences}
\end{figure}

To highlight the differences between our autoencoded, fused data product with other approaches, Figure \ref{differences} shows the differences of our consensus product minus a standard mean product (where the consensus was created by taking the mean of the four data products at each location) and a confirmatory factor analysis (CFA) product similar to that of \cite{neeley2014bayesian}.  Notably, in both cases there are regions where our consensus is higher than the others (shown in brown) while in other areas it was lower than the other methods (shown in blue).  Importantly, our product estimates a higher precipitation along the Tibetan plateau;s escarpment than in either the mean product or CFA product suggesting our autoencoding product deals with capturing the spatial features of this important area differently than the other approach.  Our autoencoding product is only lower than the mean product in the western plains.  This is likely due to the mean approach putting too much weight on APHRODITE in this area and the mean being overly sensitive to the APHRODITE outliers in this region.

In comparing our autoencoded product to either the mean or CFA approaches, note that because there is no definitive ``ground truth'' which we can use to evaluate these consensuses, we cannot make a claim that our autoencoded product is ``better'' or ``worse'' than another fused product.  To illustrate, we note that the average pseudo-$R^2$ of the autoencoded consensus product with the 4 data products is 0.815 compared to pseudo-$R^2$'s of 0.812 and 0.807 for the CFA product and mean product, respectively.  Given the similarity of pseudo-$R^2$, our autoencoded product is not substantially ``better'' than an alternative data fusion technique.  However, our autoencoder is unique in its ability to capture distinct spatial features across the various data products.

\section{Conclusions} \label{conc}

In this research, we presented a novel approach to data fusion using a spatially-varying autoencoder.  Unlike typical autoencoders, by constraining the autoencoder weights, we created a consensus data product that can by interpreted as a valid data product.  The fused consensus from our autoencoder captures different spatial features than those found using previous approaches to data fusion.  By using the Bayesian paradigm, we were able to include uncertainty quantifications on all model parameters and on the fused consensus.  

We note that there are some limitations to this research.  First, as mentioned in section \ref{app}, the MCMC algorithm required a very large number of draws and burn-in.  Even with the large number of samples and burn-in used in this application, the individual model parameters continued to slowly change as the autoencoder kept slowly learning, requiring a large number of draws for posterior inference.  Obtaining such a large number of MCMC samples is computationally intensive but perhaps using the approach of \citet{polson2017deep} or some form of Hamiltonian Monte Carlo that incorporates gradients could achieve better mixing of the Markov chain.

As written, the autoencoder assumes that each data product is available on the same spatial grid.  However, as is common for data fusion, various climate data products might have varying resolution and precision associated with their measurements.  An interesting and relevant extension of this work would consider situations where the data products are on different spatial grids and of varying fidelity.

Our autoencoder above did not explicitly account for spatial correlation in the different data products.  Rather, we assumed that the underlying output from the autoencoder sufficiently captured this spatial correlation.  Given that the autoencoder inputs are the data products themselves, assuming independence was likely approximately correct for this application.  However, further work could be done to see if accounting for spatial correlation in the output layer improves model fit and the overall estimate consensus.  However, we note that incorporating such spatial correlation into the autoencoder would likely massively increase the computation needed to fit the model.

Beyond spatial correlation, the HMA data includes nearly 20 years of monthly data.  While we successively applied the autoencoder to each month independently (see \url{https://github.com/jajcelloplayer/Weather-Autoencoder}), accounting for temporal correlation between months through, for example, a recurrent autoencoder, may also be of interest.  This is currently an area of active research.

This material is based upon work supported by the National Aeronautics and Space Administration under Grant/Contract/Agreement No.  10053957-01.


\bibliographystyle{imsart-nameyear.bst}

\bibliography{AEDF}       

\end{document}